\documentstyle[11pt,newpasp,twoside,epsf]{article}
\def\edcomment#1{\iffalse\marginpar{\raggedright\sl#1\/}\else\relax\fi}
\reversemarginpar

\newcommand{\racc}{r_{\mathrm{acc}}}
\newcommand{\tw}{t_{\mathrm{w}}}
\newcommand{\Mw}{M_{\mathrm{w}}}
\newcommand{\resc}{\left.r_{\mathrm{esc}}\right.}
\newcommand{\rph}{\left.r_{\mathrm{ph}}\right.}
\newcommand{\Tph}{\left.k T_{\mathrm{ph}}\right.}
\newcommand{\Lph}{\left.L_{\mathrm{ph}}\right.}
\newcommand{\LIS}{\left.L_{\mathrm{IS}}\right.}
\newcommand{\Msun}{\mathrm{M}_{\sun}}
\newcommand{\fgamma}{\left.f_{\gamma}\right.}

\newcommand{\ris}{r_{\mathrm{IS}}}

\newcommand{\IBand}{{\mathcal{I}}_{\mathrm{Band}}}
\newcommand{\IPlanck}{{\mathcal{I}}_{\mathrm{Planck}}}

\newcommand{\Epeak}{E_{\mathrm{p}}}

\newcommand{\fint}{\lambda} 

\begin{document}
\title{The expected photospheric emission of GRBs in the internal shock model}
 \author{Fr\'{e}d\'{e}ric Daigne and Robert Mochkovitch}
\affil{Institut d'Astrophysique de Paris, 98 bis boulevard Arago,\\
75014 Paris, France.}

\begin{abstract}
The prompt emission of gamma-ray bursts (hereafter GRBs) probably comes from a highly relativistic wind which converts its kinetic energy into radiation via the formation of shocks within the wind itself. Such "internal shocks" can occur if the wind is generated with a highly non uniform distribution of the Lorentz factor $\Gamma$.
Taking into account such a variable distribution of $\Gamma$, we estimate the expected thermal emission of the relativistic wind when it becomes transparent. We compare this emission (temporal profile + spectrum) to the emission produced by the internal shocks. In most cases we predict a rather bright thermal emission that could easily be detected. This favors acceleration mechanisms for the wind where the main energy reservoir is under magnetic rather than thermal form. Such scenarios can produce thermal X-ray precursors comparable to those observed by GINGA and WATCH/GRANAT.
\end{abstract}

\section{The photosphere of a relativistic wind}
We do not discuss here the nature of the source responsible for the initial energy release leading to the GRB. 
We assume that a relativistic wind has emerged at radius $\racc$
 with a variable distribution of the Lorentz factor $\Gamma(t)$ 
(with $\bar{\Gamma}\ga 100$). The energy and mass injection rates are $\dot{E}(t)$ and  $\dot{M}(t)=\dot{E}(t)/\Gamma(t)c^{2}$. The wind production lasts for a total duration $\tw$ (total ejected mass : $\Mw$).
At $r=\racc$, the wind is still optically thick, whereas at the radius $\ris$ where the internal shock phase starts, it has to be already transparent, as indicated by the dominant non-thermal nature of the observed GRB spectra.
The radius at which each layer of the relativistic wind becomes transparent 
can be estimated (see Daigne \& Mochkovitch 2002) : photons
 emitted by a layer at $r$ will successively cross all the layers emitted earlier by the source before they can escape from the relativistic wind at radius $\resc(r)$ : as $\Gamma \gg 1$, we usually have $\resc(r)\gg r$. The optical depth of a layer at radius $r$ is approximatively given by
$\tau(r) \simeq {\kappa \dot{E}} / {8\pi c^{3} \Gamma^{2} r}$,
where the opacity $\kappa$ is mainly due to the ambient electrons (Thomson opacity). Therefore the photospheric radius ($\tau\simeq 1$) of the layer is given by : 
$\rph \simeq {\kappa\dot{E}} / {8\pi c^{3}\Gamma^{2}}$.
Fig.~1 shows this radius for an example where $\dot{E}=10^{52}\ \mathrm{erg.s^{-1}}$, $\tw=10\ \mathrm{s}$ and the Lorentz factor is given in the same figure.
\begin{figure}
\plottwo{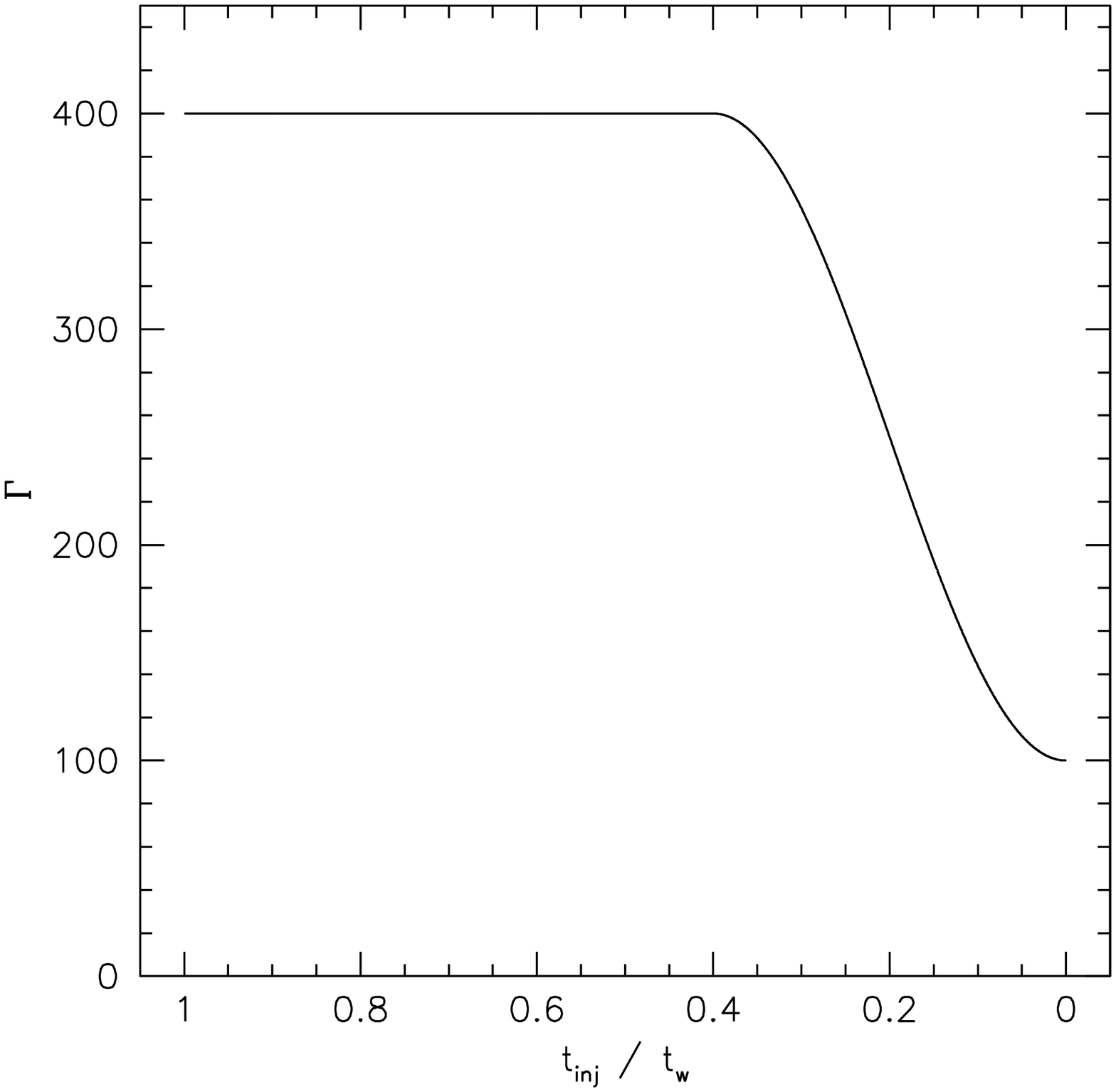}{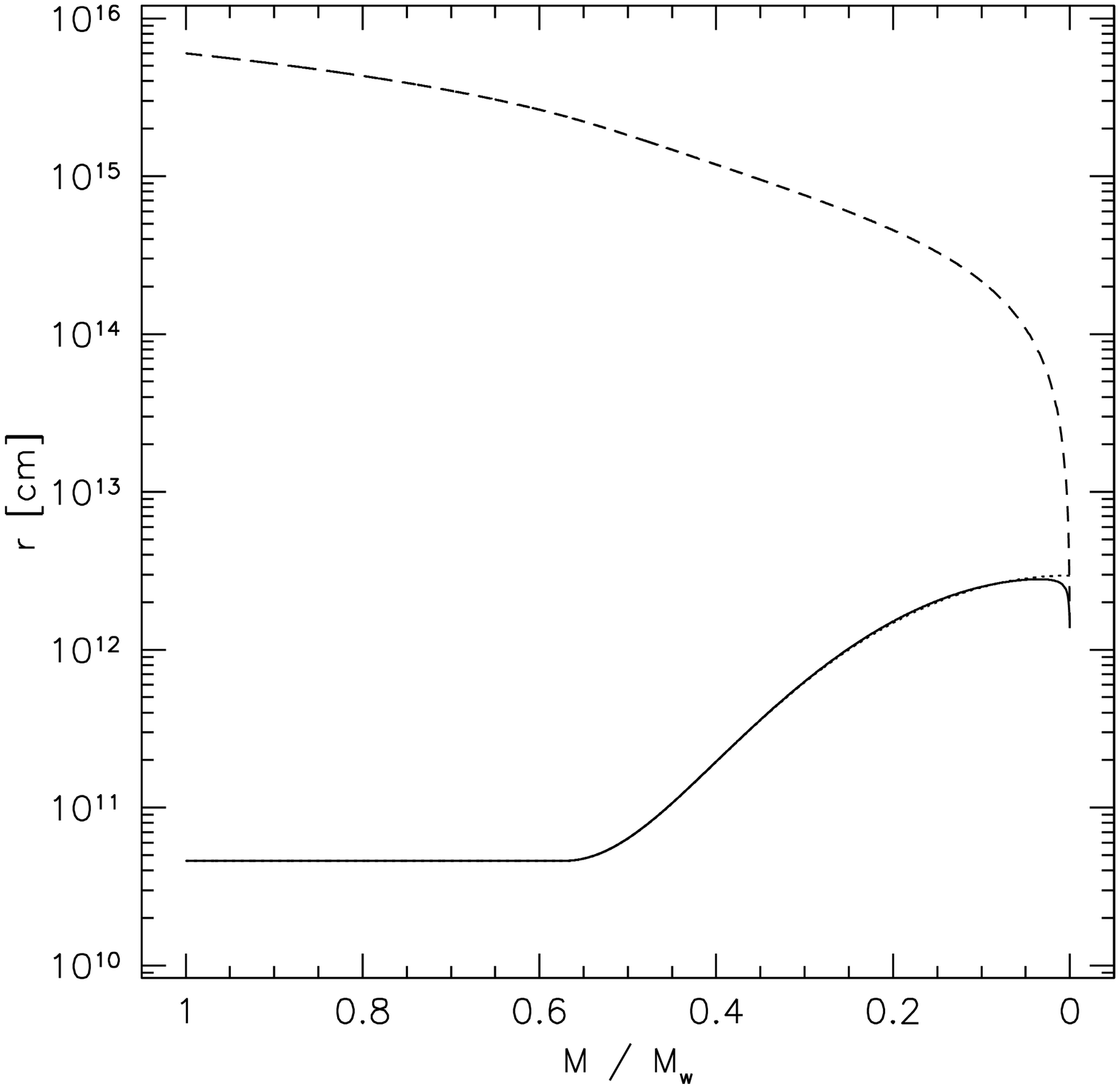}
\vspace*{-0.45cm}

\caption{\textit{Left :} initial distribution of the Lorentz factor used in our example. 
\textit{Right :} the photospheric radius of each layer (solid line) and the corresponding radius at which photons escape (dotted line) as a function of a mass coordinate in the wind (solid line).}
\end{figure}
\section{Time profile and spectrum of the photospheric emission}
The photospheric luminosity and temperature of each layer can be estimated in the context of the fireball model (see e.g. Piran 1999) :
\begin{equation}
\Lph \simeq \dot{E}\left(\frac{\rph}{\racc}\right)^{-2/3}\ {\mathrm{and}}\ 
\Tph \simeq kT^{0}\left(\frac{\rph}{\racc}\right)^{-2/3}\ ,
\end{equation}
where the initial fireball temperature is given by $kT^{0}=1.3\ \dot{E}_{52}^{1/4} \mu_{1}^{-1/2}\ \mathrm{MeV}$ ($\mu_{1}$ is the mass of the central black hole in units of $10\ \Msun$). The emitted photons will closely follow a Planck distribution, so that the phostopheric emission leads to a spectrum and a count rate in an energy band $[E_{1},E_{2}]$ :
\begin{equation}
\frac{dn^{\mathrm{ph}}(E)}{dEdt} = \frac{\left(1+z\right)\Lph}{\IPlanck\left(\Tph\right)^{2}} \times \frac{x^{2}}{\exp{x}-1}\ \mathrm{and}\
C^{\mathrm{ph}}_{12} = \frac{\Lph}{\Tph}\ \frac{1}{\IPlanck}\int_{x_{1}}^{x_{2}}\frac{x^{2}dx}{\exp{x}-1}\ ,
\end{equation}
where $x=(1+z)E/\Tph$, $x_{1,2}=(1+z)E_{1,2}/\Tph$ and $\IPlanck=\pi^{4}/15$. 

\section{Comparison with the emission from internal shocks}
The internal shock luminosity is assumed to be $\LIS=\fgamma \dot{E}$ 
where $\fgamma$ is the efficiency for the conversion of kinetic energy into $\gamma$-rays by internal shocks.
This leads to a spectrum and a count rate in energy band $[E_{1},E_{2}]$ given by
\begin{equation}
\frac{dn^{\mathrm{IS}}(E)}{dEdt} =  \frac{\LIS}{\left(1+z\right)\Epeak^{2}\IBand} {\mathcal{B}}\left(\frac{E}{\Epeak}\right)\ \mathrm{and}\ 
C^{\mathrm{IS}}_{12} = 
\frac{\LIS}{\left(1+z\right)\Epeak} \frac{1}{\IBand}
\int_{x_{1}}^{x_{2}} {\mathcal{B}}(x)dx\, 
\end{equation}

\noindent where $\Epeak$ is the observed peak energy of the internal shock spectrum ($x_{1,2}= E_{1,2} / \Epeak$)  and ${\mathcal{B}}$ is the usual Band spectral shape (with $\int_{0}^{+\infty}x{\mathcal{B}}(x)dx=\IBand$), characterized by a low and high-energy slope $\alpha$ and $\beta$ (Band et al. 1993). Fig.~2 shows the spectrum and the time profile obtained in the example of fig.~1 including both the internal shock and the photospheric emission (with $\alpha=-1.0$, $\beta=-2.25$ and $z=1$). The internal shock emission is computed with the model developped by Daigne \& Mochkovitch (1998). The global observed peak energy of the non-thermal photons is 200 keV.
\begin{figure}
\plottwo{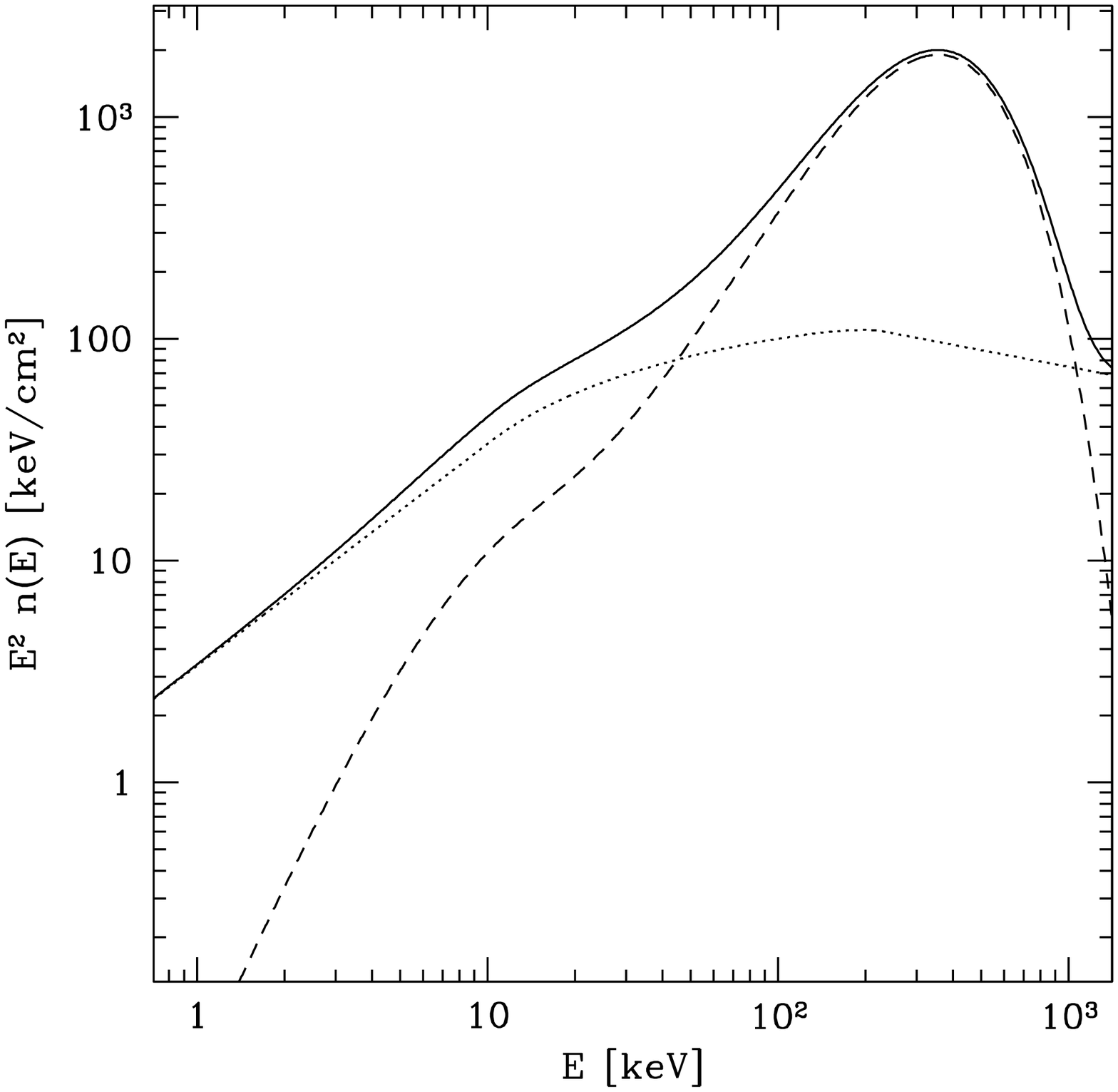}{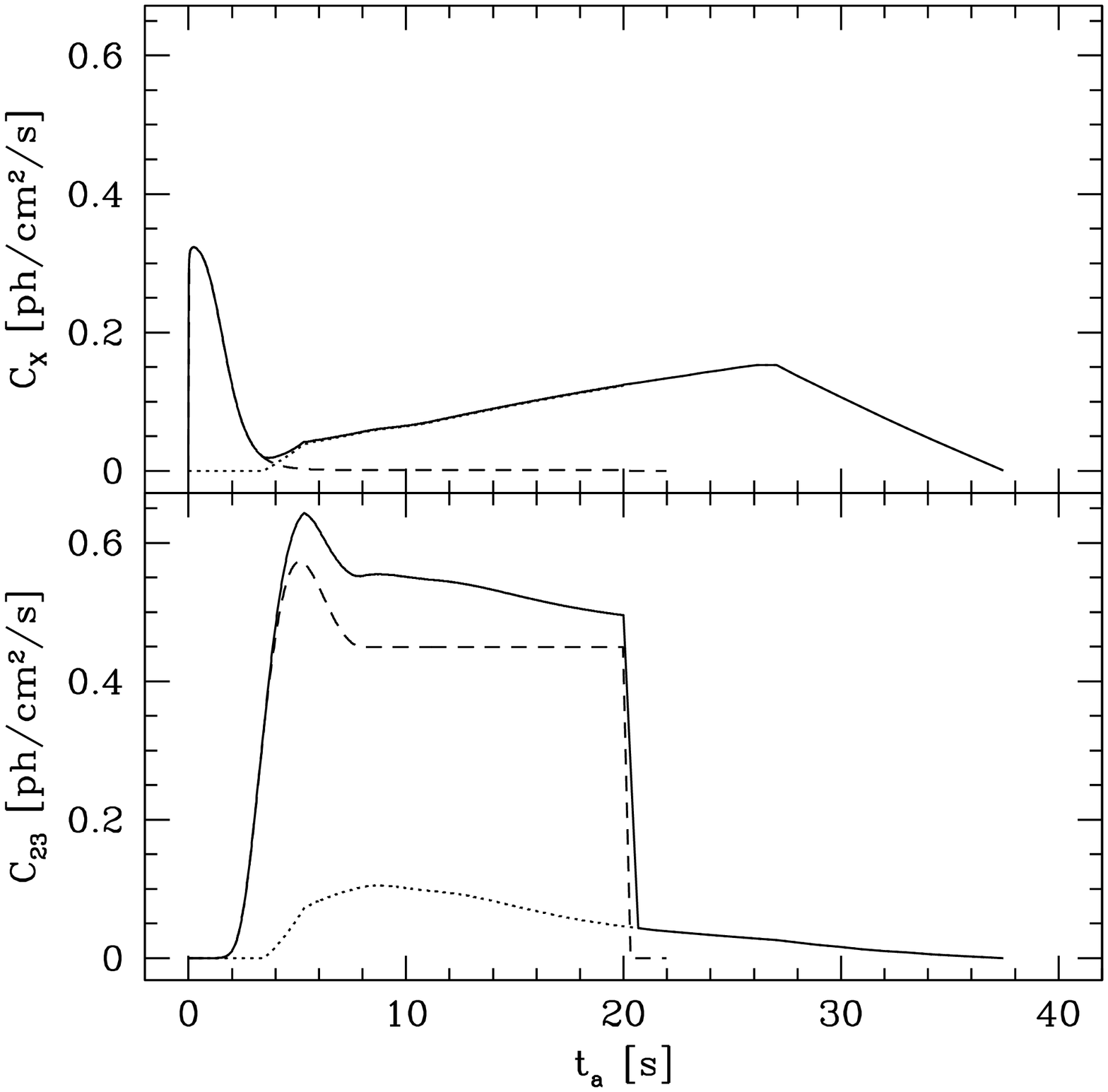}
\vspace*{-0.45cm}

\caption{Spectrum (left) and time profile (right) of the example presented in fig.~1. The time profiles are plotted in the 3.5-8.5 keV (top) and 50-300 keV (bottom) bands. The three curves correspond to the thermal photospheric emission (dashed line), the non-thermal internal shock emission (thin solid line) and the total (thick solid line).}
\vspace*{-0.5cm}

\label{fig:resultat1}
\end{figure}
\section{Discussion}
The ratio of the photospheric over internal shock count rates is given by\,:
\begin{equation}
R_{12} = \frac{C^{\mathrm{ph}}_{12}}{C^{\mathrm{IS}}_{12}} =
1.6\ \fgamma_{0.1}^{-1}\dot{E}_{52}^{-1/4}\mu_{1}^{1/2}\frac{(1+z)\Epeak}{200\ \mathrm{keV}}\frac{\IBand}{\IPlanck}\frac{\int_{x_{1}}^{x_{2}}\frac{x^{2}dx}{\exp{x}-1}}{\int_{x_{1}}^{x_{2}}{\mathcal{B}}(x)dx}\ .
\end{equation}
This expression indicates that the photospheric thermal emission is usually dominant in the X- and $\gamma$-ray bands (see fig.~2) !
This is clearly in disagreement with the observations showing that the GRB spectrum is non-thermal in this spectral range.
From eq.~4, there are two possible solutions to recover GRBs dominated by the non-thermal internal shock component : (i) increase of $\fgamma$, i.e. more efficient internal shocks. However, our calculations show that even for $\fgamma$ close to unity,
the thermal component would still be easily detectable
 ; (ii) decrease of $\Lph$ and $\Tph$, i.e. a less hot and bright photosphere, compared to the prediction of the standard fireball model.
If we define $\fint$ as the fraction of the energy released by the source which is initially under thermal form, the photospheric luminosity and temperature are given by :
\begin{equation}
\Lph \simeq  \fint\dot{E}\left(\frac{\rph}{\racc}\right)^{-2/3}\ \mathrm{and}\ 
\Tph \simeq \fint^{1/4}\ kT^{0}\left(\frac{\rph}{\racc}\right)^{-2/3}\ ,
\end{equation}
so that the expression of $R_{12}$ in eq.~4 has now to be multiplied by a factor $\fint^{3/4}$. In the standard fireball model $\fint=1$ but much smaller values can be expected if a large fraction of the energy realeased by the source is initially under magnetic form. Fig.~3 shows exactly the same example as fig.~1, the only difference being that the photospheric emission has been computed with $\fint=0.01$. A non-thermal burst is recovered in agreement with the observations. A small thermal precursor is still present in X-rays. It can be more intense for intermediate values of $\fint$, so that the photospheric emission predicted in this case could explain  the X-ray precursors observed 
in a few GRBs.
\begin{figure}
\plottwo{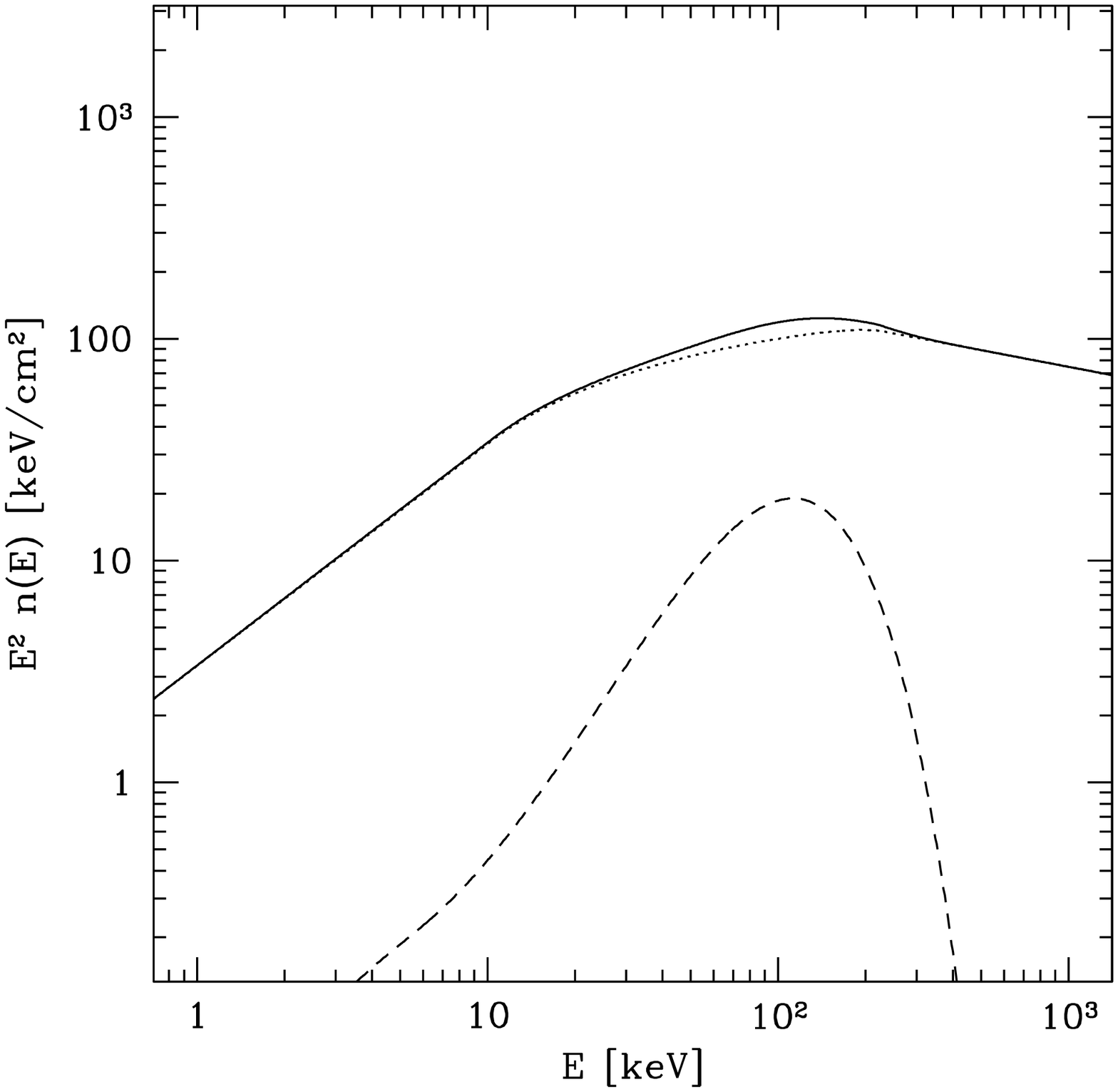}{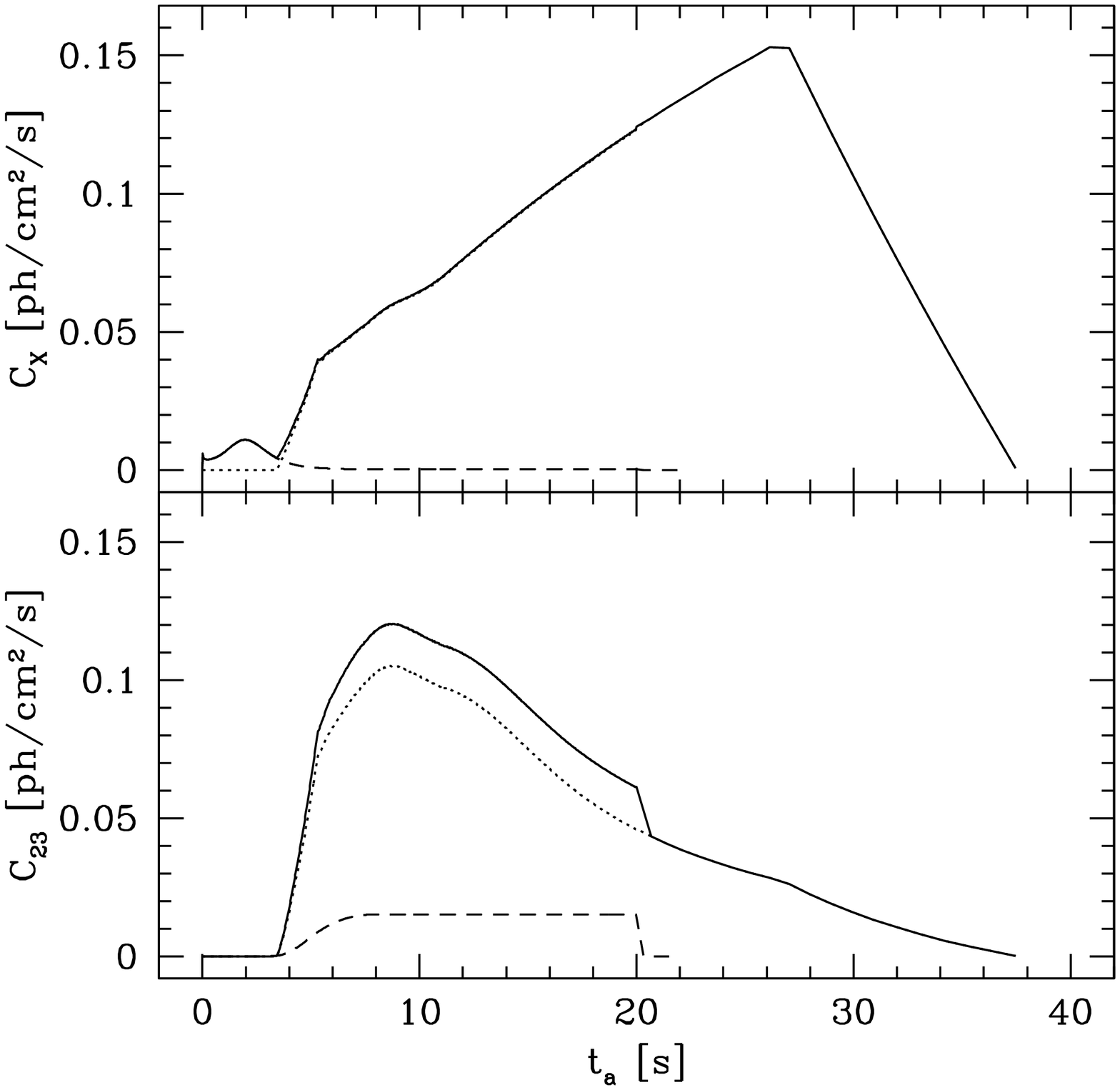}
\vspace*{-0.45cm}

\caption{Same example as in fig.~2 but with a less hot and luminous photosphere computed with $\fint=0.01$ (see text).}
\vspace*{-0.5cm}

\end{figure}

\section{Conclusion}
We have computed the photospheric thermal emission of a relativistic wind with a highly
variable initial distribution of Lorentz factor. We find that : (1) The photosphere predicted in the standard fireball model for GRBs is usually too hot and luminous. The thermal photospheric emission dominates the non-thermal internal shock emission, even for large internal shock efficiencies. This is in disagreement with the observations. (2) A less luminous photosphere is
expected if the energy released by the source is initially
injected mainly under magnetic rather than thermal form.
In this case, we recover GRB
spectra and temporal profiles dominated by the non-thermal
internal shock emission.
(3) We then obtain X-ray precursors which may correspond
to the precursor activity observed in a few GRBs 
by GINGA
and WATCH/GRANAT. These resultats are presented in more details in Daigne \& Mochkovitch (2002).

\end{document}